# The extremely low metallicity star SDSS J102915+172927: a subgiant scenario

J. MacDonald[1], T.M. Lawlor[2], N. Anilmis[1] and N.F. Rufo[2]

[1]University of Delaware, Department of Physics and Astronomy, Newark, DE 19716, USA
[2]Pennsylvania State University, Brandywine Campus, Department of Physics, Media, PA 19063 USA

**ABSTRACT**

Spectroscopic analysis of the Galactic halo star SDSS J102915+172927 has shown it to have a very low heavy element abundance, $Z < 7.4 \times 10^{-7}$, with [Fe/H] = –4.89 ± 0.10 and an upper limit on the C abundance of [C/H] < –4.5. The low C/Fe ratio distinguishes this object from most other extremely metal poor stars. The effective temperature and surface gravity have been determined to be $T_{eff}$ = 5811 ± 150 K and log $g$ = 4.0 ± 0.5. The surface gravity estimate is problematical in that it places the star between the main sequence and the subgiants in the Hertzsprung-Russell diagram. If it is assumed that the star is on the main sequence, its mass and are estimated to be $M$ = 0.72 ± 0.06 M$_\odot$ and $L$ = 0.45 ± 0.10 L$_\odot$, placing it at a distance of 1.35 ± 0.16 kpc. The upper limit on the lithium abundance, $A$(Li) < 0.9, is inconsistent with the star being a dwarf, assuming that mixing is due only to convection. In this paper, we propose that SJ102915 is a sub-giant that formed with significantly higher $Z$ than currently observed, in agreement with theoretical predictions for the minimum C and/or O abundances needed for low mass star formation. In this scenario, extremely low $Z$ and low Li abundance result from gravitational settling on the main sequence followed by incomplete convective dredge-up during subgiant evolution. The observed Fe abundance requires the initial Fe abundance to be enhanced compared to C and O, which we interpret as formation of SJ102915 occurring in the vicinity of a type Ia supernova.

**1 INTRODUCTION**

Theoretical studies (Bromm & Loeb 2003; Schneider et al. 2003) indicate that low-mass stars cannot form until the mass fraction $Z$ of heavy elements in the interstellar medium has been enriched to a critical value estimated to lie in the range $1.5 \times 10^{-8}$ to $1.5 \times 10^{-6}$. Bromm & Loeb

(2003) argue that the abundances of carbon and oxygen are the crucial factor in determining whether only massive stars can form (as in stellar population III) or both massive and low-mass stars can form (as in stellar populations II and I). In this scenario, the fine structure lines of ionized carbon and neutral oxygen provide efficient cooling of the protostellar clouds in the primitive interstellar medium. Frebel, Johnson, & Bromm (2007) define an 'observer friendly' transition discriminant $D = \log_{10}\left(10^{[C/H]} + 0.3 \times 10^{[O/H]}\right)$ such that low-mass star formation is possible only if $D > -3.5$. Here the notation, [X/H] means log of the star's abundance ratio relative to the solar abundance ratio. Frebel et al (2007) make a prediction that any star with [Fe/H] $\lesssim -4$ will have enhanced C and/or O abundances. Support for this prediction comes from the discovery of stars with very low Fe abundances and relatively high C and/or O abundances such as HE 0107-5240 and HE1327-2326 (Christlieb et al. 2002).

However, the very recent discovery of SDSS J102915+172927 (hereafter SJ102915), which is not carbon enhanced casts doubt on this picture. Caffau et al. (2012, hereafter Caf12) report that SJ102915 has [Fe/H] = $-4.89 \pm 0.10$ and has no measurable enhancement of carbon or nitrogen. They estimate that $Z \lesssim 7.4 \times 10^{-7}$. Another unusual feature of SJ102915 is the complete absence of the neutral lithium resonance doublet, a feature that is all but constant in other metal poor dwarf stars (Caffau et al. 2011).

In this paper, we use our stellar evolution models to compare the observed properties of J102915 with main sequence and subgiant model predictions. In Section 2 we briefly describe our stellar evolution code, BRAHMA. In section 3, we consider models in which the effects of gravitational settling and element diffusion are neglected. We also discuss here the 'lithium problem'. In section 4, we show that including the effects of gravitational settling and element diffusion resolves the lithium problem provided that SJ102915 is a subgiant star. Our conclusions and discussion are given in section 5.

**2 THE EVOLUTION CODE**

Here we briefly describe our evolution code, BRAHMA (Mullan & MacDonald 2010; Lawlor et al. 2008; Lawlor & MacDonald 2006). The code uses a relaxation method to simultaneously solve the stellar structure equations along with adaptive mesh and composition equations for the star as a whole. OPAL opacities (Iglesias & Rogers 1996) are used for temperatures greater than 6000 K with a smooth transition to the Ferguson et al. (2005) opacities at lower temperatures. Interpolation in the opacity tables is handled by using the subroutines of Arnold Boothroyd (which are obtainable from http://www.cita.utoronto.ca/~boothroy/kappa.html) . Convective energy transport is treated by mixing length theory as described by Mihalas (1978), which is the same as that of Böhm-Vitense (1958) but modified to include a correction to radiative losses from convective elements when they are optically thin. The nuclear reaction network explicitly follows the evolution of the isotopes $^{1}$H, $^{2}$H, $^{3}$He, $^{4}$He, $^{7}$Li, $^{7}$Be, $^{12}$C, $^{13}$C, $^{14}$N, $^{16}$O, $^{20}$Ne, $^{24}$Mg, $^{28}$Si and $^{56}$Fe. All the nuclear reaction rates relevant to hydrogen burning are from Angulo et al.

(1999), except for $^{14}N(p,\gamma)^{15}O$ (Herwig & Austin 2004). The electron screening enhancement factor for a nuclear reaction is taken to be the lowest of the weak screening (Salpeter 1954), intermediate screening (Graboske et al. 1973) and strong screening factors (Itoh et al 1979; Itoh et al 1990). Composition changes due to convective mixing are treated by adding diffusion terms to the composition equations, with the diffusion coefficient consistent with mixing length theory.

For the calculations in which we include composition changes due to element diffusion and gravitational settling, the diffusion velocities for all 14 species in the nuclear reaction network are calculated by numerically solving the multicomponent flow equations derived by Burgers (1969) and summarized by Muchmore (1984). Our earlier use of this approach for modeling element diffusion processes in white dwarf stars can be found in Iben & MacDonald (1985), Iben, Fujimoto & MacDonald (1992), MacDonald, Hernanz & Jose (1998). More recently, we have used the Burger's formulation in modeling the Sun (Mullan, MacDonald & Townsend 2007).

We have not included radiative levitation in our diffusion calculations. Even at the relative low luminosities of the low mass stars considered here, radiative levitation of elements with very low abundances may be important. Seaton (1997) has addressed calculation of the radiative accelerations using OP data for a number of elements. From his figure 6, the maximum value of radiative acceleration on Fe at temperature $2\times10^5$ K at 1/100 the solar abundance is determined to be about 4000 times the radiative acceleration of free electrons. The radiative acceleration of Fe will be larger still at lower abundances. Exploratory calculations that include radiative of Fe are presented in appendix A.

## 3 MODELS WITHOUT ELEMENT DIFFUSION OR GRAVITATIONAL SETTLING

We have evolved models of initial composition $X = 0.765$, $Y = 0.235$ and $Z = 1.86\times10^{-6}$ (which corresponds to $10^{-4}$ times the solar system heavy element abundance), for masses from $M = 0.50$ $M_\odot$ to $0.90$ $M_\odot$, in increments of $0.05$ $M_\odot$. We use our solar calibrated mixing length ratio $\alpha = 1.70$. All models are evolved from the pre-main sequence to the age of the Universe ($1.37\times10^{10}$ yr) or, for the more massive models, to the onset of the helium core flash. We do not consider masses higher than $0.90$ $M_\odot$ for two reasons: 1) For more massive stars of this very low $Z$, the evolutionary paths in the log $T_{eff}$ – log $g$ diagram get further away from the observations, and 2) more massive stars reach the helium flash in less than $10^{10}$ years. Since the oldest globular clusters have age > 12 Gyr and $Z \geq 0.0002$ (Jimenez et al. 1996; Salaris, degl'Innocenti & Weiss, 1997), it seems improbable that a star could form $10^{10}$ years ago with $Z \approx 10^{-6}$.

For comparison with the observational data, we give in Figure 1 evolutionary tracks in the log $T_{eff}$ – log $g$ diagram. The corresponding Hertzsprung-Russell diagram is shown in figure 2.

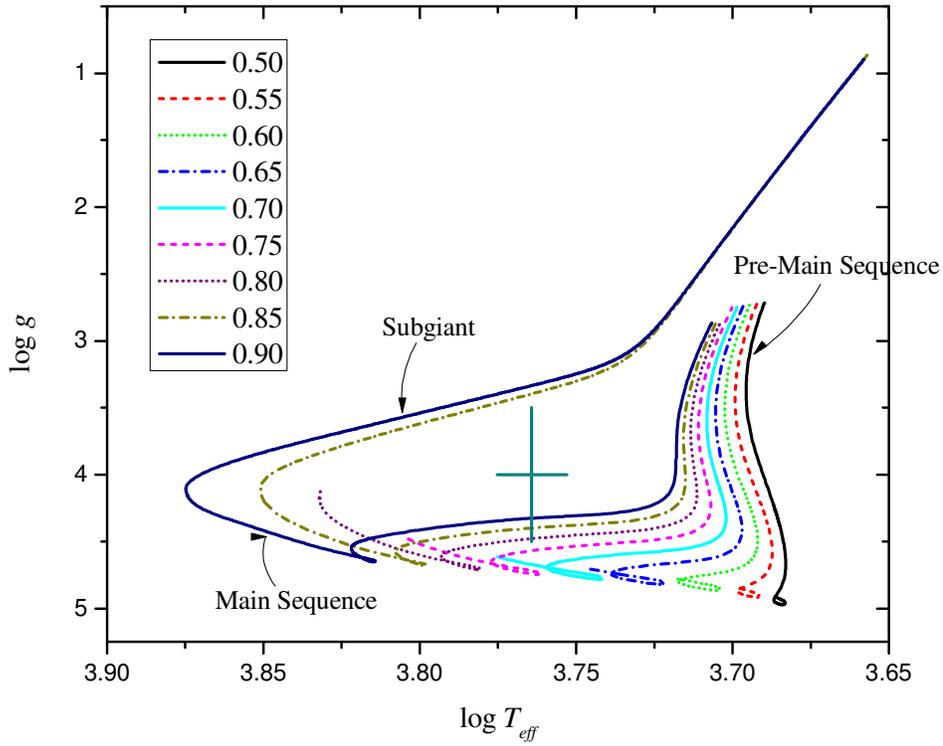

**Figure 1.** Evolutionary tracks in the log $T_{eff}$ - log $g$ diagram for models of initial composition $X = 0.765$, $Y = 0.235$ and $Z = 1.86\times10^{-6}$. The mixing length ratio is $\alpha = 1.7$. The spectroscopic determinations of Caf12 for SJ102915 are shown by error bars (dark cyan lines). The units of $T_{eff}$ and $g$ are K and cm s$^{-2}$, respectively. The legend labels the tracks by mass in solar units.

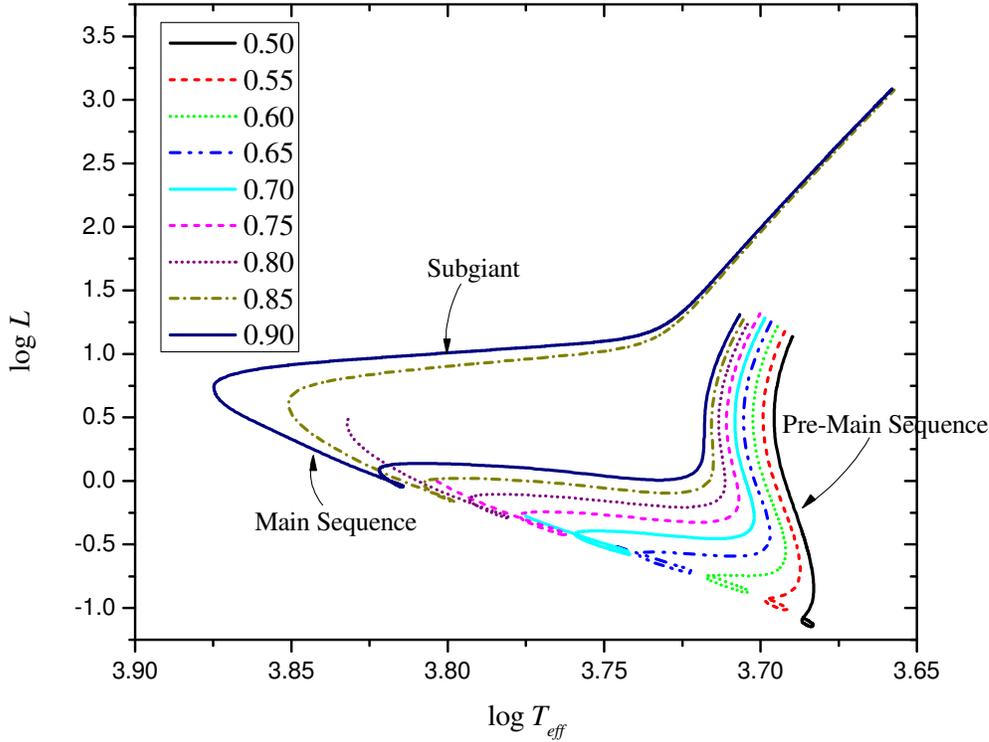

**Figure 2**. Evolutionary tracks in the Hertzsprung-Russell diagram corresponding to the tracks shown in figure 1. The units of $T_{eff}$ and $L$ are K and $L_\odot$, respectively. The legend labels the tracks by mass in solar units.

Because of the low heavy element abundance, we dismiss the possibility that SJ102915 is in the short-lived pre-main sequence phase of evolution. Figure 1 shows that main sequence models do not provide a good match to the surface gravity. However, sub-giant models are only marginally better. If we ignore the gravity estimate, inspection of figure 1 shows that the effective temperature constrains the mass to the range 0.66 – 0.78 $M_\odot$, assuming that SJ102915 is a dwarf star. If we further assume that SJ102915 is old, e.g. has an age of 12 Gyr, then its mass is constrained to be 0.67 – 0.71 $M_\odot$, in agreement with the mass found by Caf12 using unpublished Chieffi & Limongi models. As can be seen from figure 2, the corresponding luminosity range for our models is $L$ = 0.35 – 0.54 $L_\odot$ ($M_{bol}$ = 5.41 – 5.88). To determine the predicted distance, we first estimate the apparent bolometric magnitude of SJ102915 by using the transformations between the Johnson-Cousins *UBVRI* and SDSS *ugriz* systems given by Jordi, Grebel & Ammon (2006). For $g$ = 16.922 ± 0.004 and $r$ = 16.542 ± 0.004, we obtain $V$ = 16.686 ± 0.008. We determine the bolometric correction (BC) by using results from the NextGen atmosphere models (Hauschildt, Allard & Baron 1999). Using the tables given on the web site of France Allard (http://phoenix.ens-lyon.fr/Grids/NextGen/), we obtain for the main sequence model, BC = -0.26

± 0.01. After correcting for extinction, we obtain $m_{bol}$ = 16.37 ± 0.02. The resulting distance is 1.19 – 1.51 kpc.

Alternatively, if we assume that SJ102915 is a subgiant, then the spectroscopic $T_{eff}$ constrains the mass to be greater than 0.815 $M_\odot$. For an assumed age of 12 Gyr, the mass must be close to 0.845 $M_\odot$. Using a bolometric correction appropriate to a subgiant, the range in model luminosity, $L$ = 8.7 – 9.7 $L_\odot$ ($M_{bol}$ = 2.27 – 2.39), implies a distance of 6.0 – 6.4 kpc.

We note that one uncertain factor that may improve the log $g$ fit is the mixing length ratio. We have also calculated models with $\alpha$ = 1.00 but find that the fit is not improved and the log $g$ problem remains. Also our adopted helium abundance, from Peimbert, Peimbert & Ruiz (2000). is lower than more recent determinations of the primordial helium abundance, $Y_p$ = 0.2477 ± 0.0029 (Peimbert, Luridiana & Peimbert 2007), $Y_p$ = 0.2565 ± 0.0060 (Izotov & Thuan 2010). Models calculated with $Y$ = 0.250 are in general hotter at a given log $g$ than those for $Y$ = 0.235. At a given $T_{eff}$, the trend with increasing $Y$ is that log $g$ increases on the main sequence but decreases on the subgiant branch. Hence a higher initial $Y$ value makes the fit between observations and theory poorer in both the main sequence and subgiant scenarios. The shift in log $g$ on the subgiant branch is of order 0.04, which is small compared to the uncertainty in the observationally determined value of log $g$.

### 3.1 The lithium problem

In terms of $A$(Li) = 12 + log($N$(Li)/$N$(H)), Caf12 find an upper limit on the lithium abundance of $A$(Li) < 0.9, far below the Spite 'plateau' value of $A$(Li) ≈ 2.2 (Spite & Spite 1982) and the big bang nucleosynthesis value, $A$(Li) = 2.72 ± 0.05 (Cyburt et al. 2008). Figure 3 shows the time evolution of the $^7$Li abundance for our models of stars in the mass range 0.50 – 0.80 $M_\odot$. We see that the observed lithium depletion is inconsistent with the range in mass determined from $T_{eff}$ under the assumption that SJ102915 is a dwarf star.

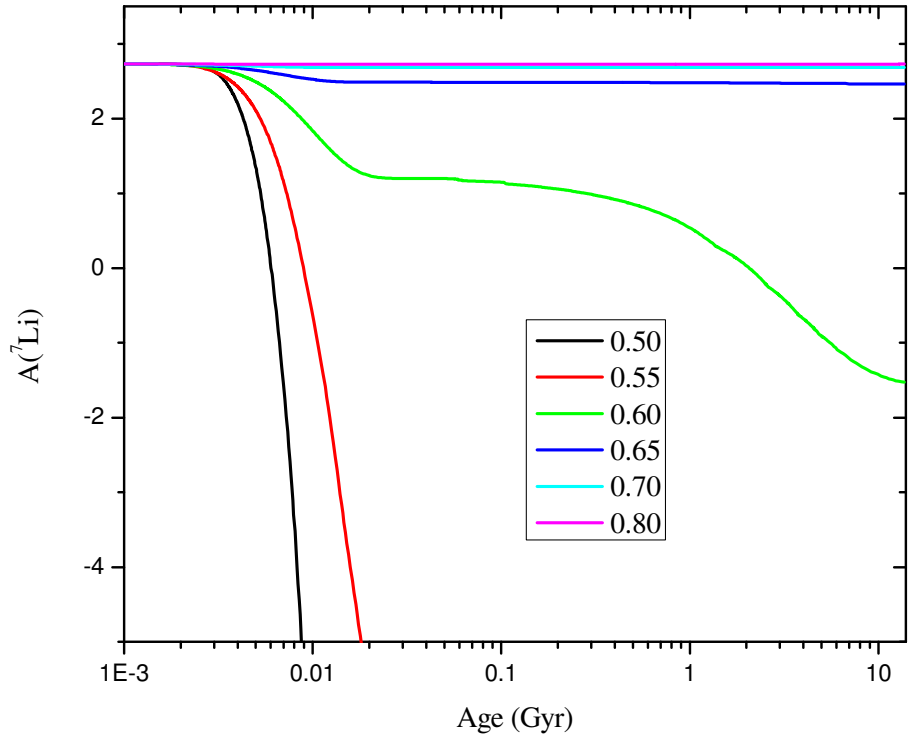

**Figure 3.** Evolution of the surface lithium abundance for model stars of mass 0.50 – 0.80 M$_\odot$.

Figure 4 shows the lithium abundance evolution for models that reach the sub-giant phase of evolution. Although some reduction of the lithium abundance does occur during the dredge-up phase, it mainly occurs after $T_{eff}$ is below the lower limit found for SJ102915. Even then the lithium abundance is higher than the spectroscopic upper level.

    From this we conclude that standard stellar evolution models for SJ102915 that only include convective mixing are inconsistent with the observed $^7$Li abundance limit. This suggests

that other physical processes beyond standard convection are required.

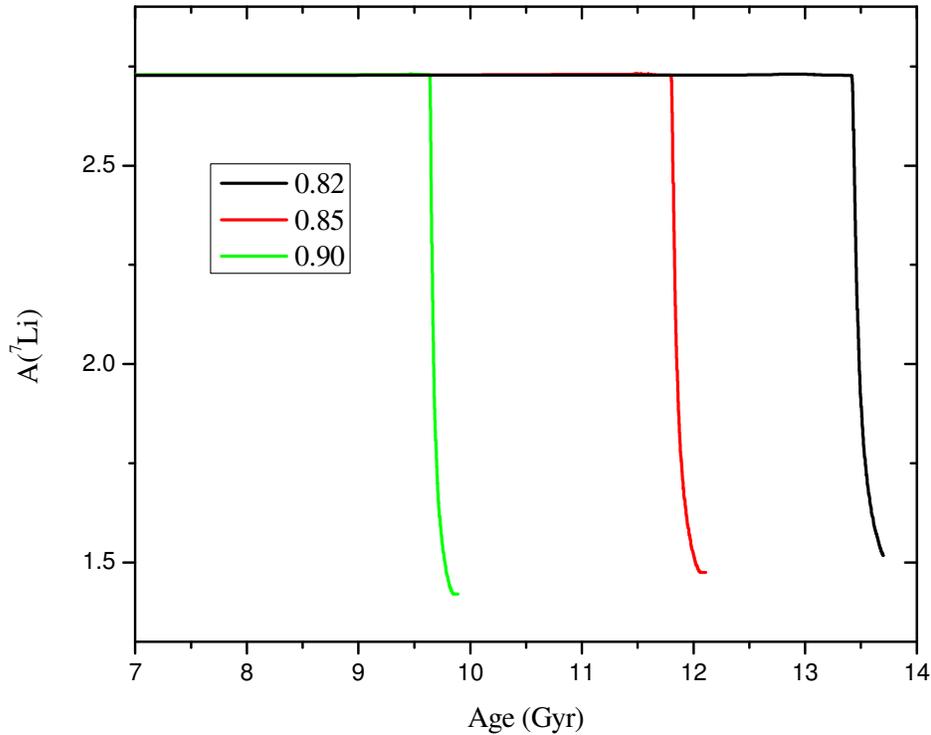

**Figure 4.** Evolution of the surface lithium abundance for model stars that reach the subgiant phase of evolution.

## 4 MODELS WITH ELEMENT DIFFUSION AND GRAVITATIONAL SETTLING

Models of low mass main sequence stars with extremely low heavy element abundance have higher surface gravity than models of population I stars of the same mass. As a consequence they are hotter and have shallower surface convection zones. The higher gravity and shallower convection leads to faster settling of heavy elements out of the surface layers, and so gravitational settling will be able to modify the surface composition for main sequence stars of lower mass than for Pop I composition stars (assuming that there are no competing mechanisms other than convective mixing). We illustrate this point by showing in figure 5 the evolution of the photospheric carbon abundance for 0.8 $M_\odot$ models of different initial $Z$.

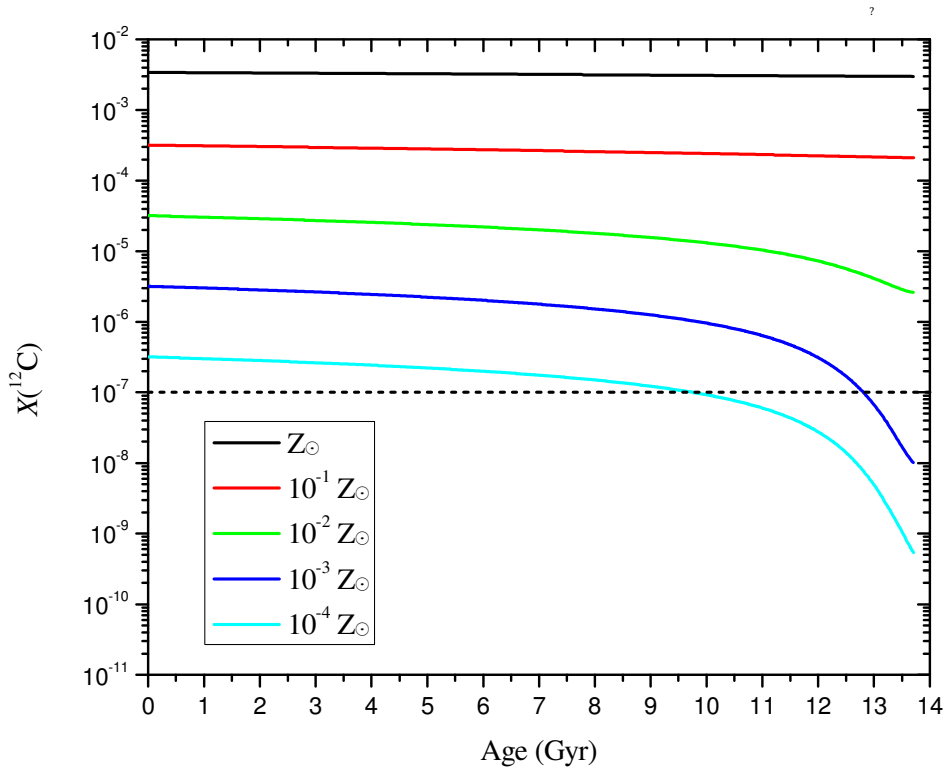

**Figure 5.** The evolution of the photospheric carbon abundance for 0.8 M$_\odot$ models of different initial heavy element abundance. The horizontal broken line shows the upper limit on the carbon abundance found by Caffau et al. (2012).

We see that gravitational settling can reduce the $^{12}$C abundance below the upper limit found by Caf12 for initial abundances as high as $10^{-3}$ Z$_\odot$. The effects of gravitational settling are even more pronounced at higher mass, but become negligible at lower mass because of the deeper surface convection zones. Since the subgiant model requires the mass to be greater than 0.8 M$_\odot$, we consider the possibility that the initial heavy element abundances of SJ102915 were greater than what is observed today, and gravitational settling is responsible for reducing them to the observed levels.

    We first explore the constraints imposed by the $^7$Li upper limit. In figure 6, we show how the Li abundance changes with $T_{eff}$ for 0.85 M$_\odot$ models of initial heavy element abundance $10^{-4}$ and $10^{-3}$ Z$_\odot$ for mixing length ratios $\alpha = 1.3$ and 1.7. The general trend of the photospheric $^7$Li abundance is that it decreases to very low values during the main sequence phase of evolution due to gravitational settling. The $^7$Li abundance recovers to almost its initial value during the subgiant phase due to convective dredge-up, before it becomes slightly depleted during the red giant phase due to proton captures at the base of the surface convection zone. It is clear from

figure 6 that the lithium abundance on the subgiant phase is sensitive to the mixing length ratio. For $\alpha = 1.7$, the model $^7$Li abundances at $T_{eff}$ values consistent with those inferred for SJ102915 are higher than the upper limit found by Caf12. In contrast, the model $^7$Li abundances for $\alpha = 1.3$ are consistent with the upper limit. Hence subgiant models that are consistent with the Li abundance can be found by reducing the mixing length ratio below the solar calibrated value. For the $\alpha = 1.3$ models, the log $g$ values on the part of the subgiant phase within the observe $T_{eff}$ limits are between 3.5 and 3.6, which are also consistent with observations.

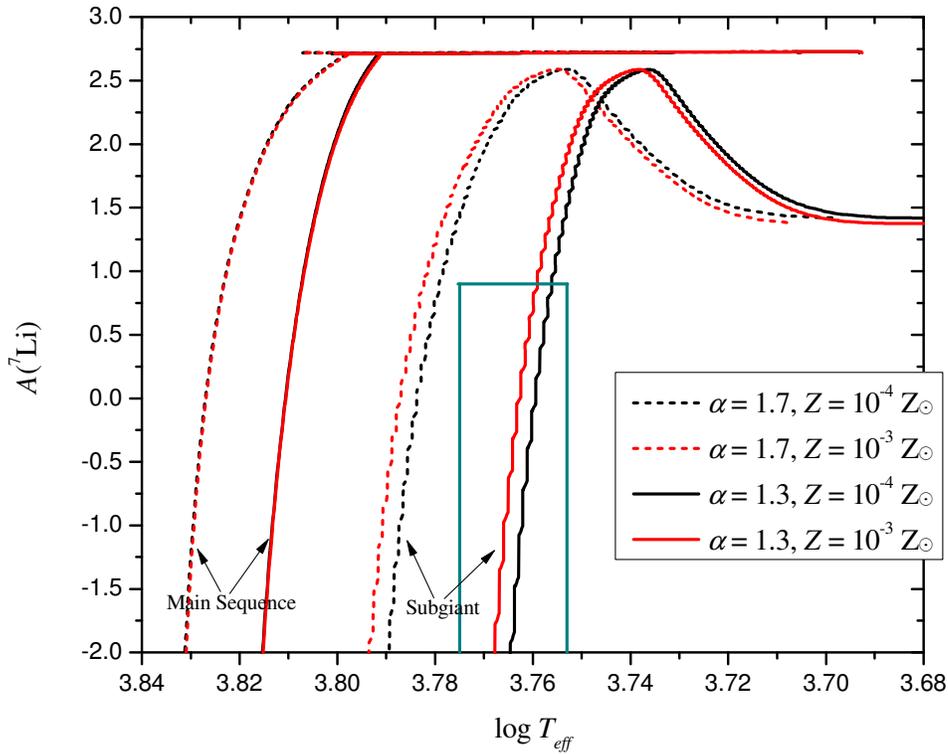

**Figure 6.** Variation of the surface $^7$Li abundance with $T_{eff}$ for 0.85 $M_\odot$ models of initial heavy element abundances $10^{-4}$ and $10^{-3}$ $Z_\odot$ for mixing length ratios $\alpha = 1.3$ and 1.7. The dark cyan box shows the bounds from the spectroscopically determined limits on $T_{eff}$. and the upper limit on the Li abundance.

We now consider the constraints imposed by the other abundance determinations. Our approach is to determine the depletion with time or $T_{eff}$ of a species relative to its initial abundance. We then use the observed abundances of SJ102915 to constrain its initial abundances. Figure 7 shows the degree of depletion of the heavy elements for a 0.85 $M_\odot$ model of initial heavy element abundance $10^{-3}$ $Z_\odot$ and mixing length ratio $\alpha = 1.3$. We see that the degree of depletion during the subgiant phase for the spectroscopically determined temperature range is sensitive to the value of $T_{eff}$. We also see that the depletions of Fe, Mg, and Si are greater than for C, N, and

O. For this particular model, the Li abundance upper limit requires that $T_{eff} > 5740$ K. The degree of depletions for Si and Fe require that their minimum initial abundances consistent with the observed abundance must be [Si/H] = – 1.49 and [Fe/H] = – 0.98. Required lower limits on initial abundances are given in Table 1.

**Table 1.** Minimum initial abundances consistent with observed abundances in the subgiant scenario

| Element | Abundance [X/H] |
|---|---|
| C | *-2.4*[a] |
| N | *-3.1*[a] |
| Mg | -2.69 |
| Si | -1.49 |
| Fe | -0.98 |

[a]values in italics assume that current abundances are equal to the upper bounds given by Caffau et al. (2012)

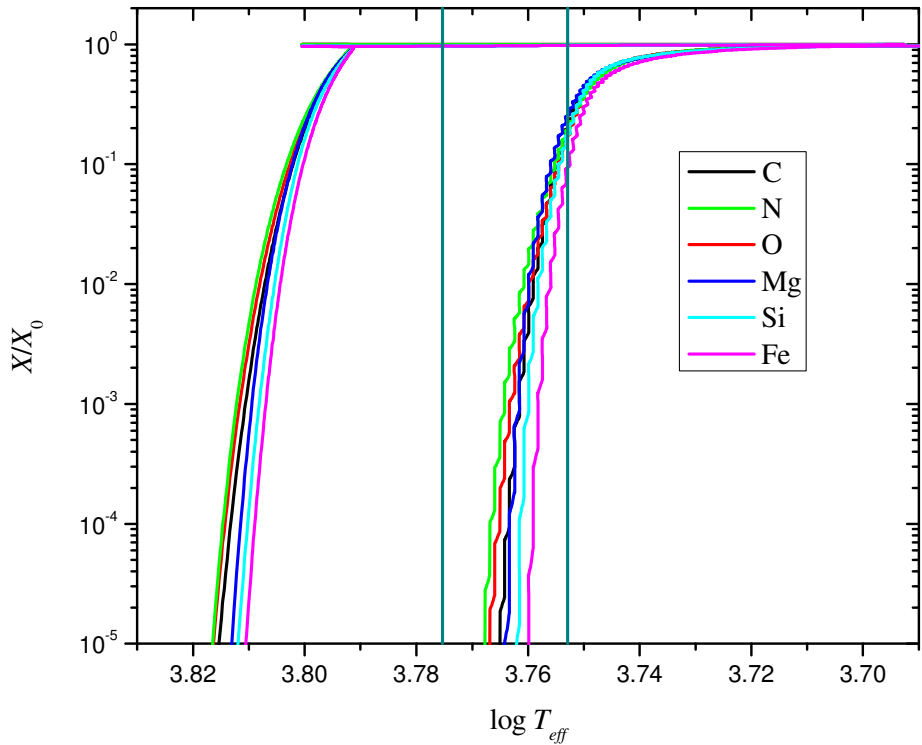

**Figure 7.** Degree of depletion of the heavy elements for a 0.85 $M_\odot$ model of initial heavy element abundance $10^{-3}$ $Z_\odot$ and mixing length ratio $\alpha = 1.3$.

The results in table 1 indicate that in the subgiant scenario the initial silicon and iron abundance must be enhanced relative to the other initial abundances. For elements other than Si and Fe, their abundances are consistent with initial values of about 1/300 solar.

## 5 CONCLUSIONS AND DISCUSSION

The Galactic halo star SDSS J102915+172927 is an unusual object in that it has a very low Z (< $7.4 \times 10^{-7}$), and a low C/Fe ratio that distinguishes it from most other extremely metal poor stars. Caffau et al. (2012, Caf12) determined effective temperature and surface gravity $T_{eff}$ = 5811 ± 150 K and log $g$ = 4.0 ± 0.5. The surface gravity estimate is problematical in that it places the star between the main sequence and the subgiants in the Hertzsprung-Russell diagram. If we assume that the star is on the main sequence, we estimate from $T_{eff}$ alone that its mass and luminosity to be $M$ = 0.72 ± 0.06 $M_{\odot}$ and $L$ = 0.45 ± 0.10 $L_{\odot}$, placing it at a distance of 1.35 ± 0.16 kpc. However the upper limit on the lithium abundance, $A$(Li) < 0.9, is inconsistent with the star being a dwarf, assuming that mixing is due only to convection. We, therefore, propose an alternative scenario in which SDSS J102915+172927 is currently in the subgiant phase of evolution. To reach the subgiant phase in the age of the Universe requires a mass > 0.815 $M_{\odot}$. Since stars of this mass with low heavy element abundance have higher surface gravity and shallower surface convection zones compared to population I stars of the same mass, gravitational settling of heavy elements has a significantly larger effect on surface abundances. We show that, in the absence of mixing processes other than convection, gravitational settling reduces the surface lithium and heavy element abundances to essentially zero during the main sequence phase of evolution. Convective dredge-up during the subgiant phase restores the abundances to about their initial values. We have shown that the effective temperature at which this occurs is sensitive to adopted value for the mixing length ratio, $\alpha$.

In this scenario, the observed lithium depletion is a result of SDSS J102915 being at an evolutionary stage in which convective dredge-up has not yet completed. To obtain consistency between the constraints set by the upper limit on the lithium abundance and the range in $T_{eff}$, we find that $\alpha$ must be less than about 1.5, which is lower than our solar calibrated value of $\alpha$ = 1.7.

We further find that the initial abundances required to give the current epoch abundances are broadly consistent with [M/H] = –2.5, with the exceptions of silicon and iron which we find must have a significantly higher initial abundances of [Si/H] ~ –1.5 and [Fe/H] ~ –1.

In the subgiant scenario, the spectroscopic $T_{eff}$ constrains the mass to be greater than 0.815 $M_{\odot}$. If we assume an age of 12 Gyr, the mass must be close to 0.845 $M_{\odot}$. The corresponding luminosity range, $L$ = 9.2 ± 0.5 $L_{\odot}$, implies a distance of 6.2 ± 0.2 kpc. Caf12 discuss the limits placed on the distance by interstellar absorptions of the Na I D-line doublet at 589.0 nm and the Ca II-K and H lines at 393.3 and 396.8 nm. By using Na I as a tracer of neutral hydrogen, they infer a hydrogen column density similar to that directly measured in $\rho$ Leo, which has a Hipparcos parallax of 0.60 ± 0.18 mas, corresponding to a distance of 1.3 – 2.4 kpc. Caf12 interpret this as setting a lower limit to the distance of SDSS J102915+172927 of 1.3 kpc, in good agreement with the distance from main sequence fitting. However, Caf12 also point out that the Na I column density is similar to that observed towards $\eta$ Leo, which has a Hipparcos parallax of 2.57 ± 0.16 mas, placing it at a distance of 0.37 – 0.41 kpc. Hence, we interpret the interstellar absorption measurements as indicating that the absorbing material is mainly in the Galactic plane, and limits the distance of SDSS J102915 only to being greater than ~ 0.4 kpc. This is also consistent with the larger distance required by the subgiant scenario.

Another important aspect of the subgiant scenario is that allows SJ102915 to have formed from material with [C/H] ~ -2.5 in agreement with the theoretical result of Frebel, Johnson, & Bromm (2007A) that their 'observer friendly' transition discriminant $D$ must be greater than -3.5 for low mass star formation to occur.

In terms of mass fractions, the inferred initial abundances in our subgiant scenario are $X(^{12}C)$ ~ $10^{-5}$, $X(^{16}O)$ ~ $3\times10^{-5}$, $X(^{28}Si)$ ~ $4\times10^{-5}$ and $X(^{56}Fe)$ ~ $1.5\times10^{-4}$. The ratios of these mass fractions are in good agreement with those from simulations of type Ia supernovae (Iwamoto et al. 1999) but differ significantly from the nucleosynthetic yields of population III type II supernovae models (Tominaga, Umeda & Nomoto 2007) which give carbon abundances equal to or greater than the abundances of iron-peak elements. Hence we propose that SDSS J102915 formed from ISM material of heavy element composition dominated by ejecta from a type Ia supernova, but that most other extremely metal poor stars formed from a more homogeneously mixed ISM containing material from type II supernovae.

In conclusion, we have shown that the observed abundances in the extremely metal poor star SDSS J102915+172927 are better explained by a model in which the star is in the subgiant phase of evolution rather than being a main sequence star. In our models, we included the heretofore neglected effects of gravitational settling and element diffusion which have significant impact on the surface abundances of extremely metal poor stars that are massive enough to evolve to the subgiant phase in times less than the age of the Universe. However, we have not included the radiative force on the individual elements. Exploratory calculations that include the radiative force on Fe indicate that radiative levitation may also play an important role in the evolution of the surface abundances of extremely metal poor stars.

## ACKNOWLEDGMENTS

JM thanks John Gizis for valuable discussions. We also thank an anonymous referee for many useful suggestions for improving the manuscript and for pointing out that radiative levitation of Fe may be an important effect in stars of very low heavy element abundance.

Tominaga N., Umeda H., Nomoto K., 2007, ApJ, 660, 516

**APPENDIX A: RADIATIVE LEVITATION OF FE**

In this appendix we present the results of some preliminary calculations that include radiative levitation of Fe. In these calculations only Fe is allowed to diffuse independently and all other elements are assumed to have the same diffusion velocity. To calculate the radiative acceleration of Fe we use Opacity Project (OP) routines (Seaton 2005) downloaded from http://cdsweb.u-strasbg.fr/topbase/TheOP.html. Specifically the routine accv.f is used to calculate the Rosseland mean opacity, $\kappa_R$, and the dimensionless radiative acceleration parameter, $\gamma$, for a grid of values of temperature, $T$ and electron number density, $N_e$. For simplicity, we use a mixture of H and Fe. At each $(T, N_e)$ grid point, the accv.f routine provides $\kappa_R$ and $\gamma$ for a specified range of values of an abundance multiplier, $\chi$, which in the current context is a measure of the ratio of Fe to H number densities, with $\chi = 1$ corresponding to the solar ratio. At each $(T, N_e)$ grid point, we use a Gaussian function to fit $\kappa_R$ and $\gamma$ as functions of $\chi$. We then use bilinear interpolation in the grid to evaluate $\kappa_R$ and $\gamma$.

The radiative acceleration, $g_{rad}$, of Fe is related to $\gamma$ and other quantities by

$$g_{rad} = \frac{1}{c} \frac{M}{M(\mathrm{Fe})} \gamma \kappa_R F, \tag{A1}$$

where $F$ is the radiative flux, $M$ is the mean atomic mass and $M(\mathrm{Fe})$ is the mass of an atom of Fe. To avoid reference to the opacity, we find it convenient to relate the radiative acceleration to the gradient of the radiation pressure,

$$g_{rad} = \gamma \frac{M}{M(\mathrm{Fe})} \frac{dp_{rad}}{dp} g. \tag{A2}$$

We have evolved stars of masses 0.7 and 0.85 $M_\odot$, heavy element abundance $Z = 10^{-4} Z_\odot$ and mixing length ratio $\alpha = 1.7$. The surface Fe abundance for the 0.7 $M_\odot$ model slowly declines during the main sequence phase, reaching 92% of its initial value at the time when $T_{eff}$ equals that of SDSS J102915. When radiative levitation is ignored the reduction is to 84%. We conclude that radiative levitation is likely to have only a small effect on the evolution of the elemental abundances in the main sequence scenario for SDSS J102915. The evolution of the surface Fe abundance is markedly different for the 0.85 $M_\odot$ model. During the main sequence phase, the Fe abundance increases significantly reaching a mass fraction of 0.0029 at the end of the main sequence phase. The large enhancement is due to the radiative force exceeding the force of gravity just below the convection zone during the later parts of the main sequence phase. As the star evolves on the main sequence, the mass of the convection zone decreases from an initial value of $10^{-3}$ $M_\odot$ to $10^{-8}$ $M_\odot$ at the end of the main sequence. Once the radiative force exceeds the force of gravity just below the convection zone, all of the Fe is 'trapped' in the convection which leads to the large surface Fe abundance increase. After the model leaves the main sequence, the convection zone mass increases and the surface Fe abundance decreases. When the model $T_{eff}$ on the subgiant branch equals that of SDSS J102915, the Fe abundance is about twice the initial abundance. Although this model is incomplete, e.g. it does not include the radiative force on other elements nor does it allow the other elements to diffuse independently, it does indicate that radiative levitation could be an important physical process in determining the evolution of the surface abundances of extremely metal poor stars.